# Enhancement of Particle Image Velocimetry Images

**S.Anand, R.Poovitha, K.Nikhila**
Department of Electronics and Communication Engineering,
Mepco Schlenk Engineering College, Sivakasi, Tamil Nadu, India

*Abstract----* **Particle Image Velocimetry (PIV) is a method to visualize the flows and quantitatively map the flows. It is used to obtain the instantaneous velocity, vorticity, divergence, shear in fluids, etc. Laser Doppler velocimetry and hot wire anemometry are other techniques that are used to measure the flow. The main difference between PIV and those techniques is that PIV produces two-dimensional or three-dimensional vector fields. PIV uses two sequences of images (frames) and the frames are split into a large number of interrogation windows. The displacement vector is calculated for each window with the help of auto-correlation or cross-correlation techniques. Missing data during these processes leads to inaccurate results. To overcome this, preprocessing techniques such as histogram equalization and intensity clipping are used to enhance the PIV images to obtain accurate results.**

## I. INTRODUCTION

Particle image velocimetry (PIV) is a standard process to measure the instantaneous velocities of fluid flow [1] – [14]. It is used to measure macro-scale fluid flows. It has applications in biomedical, aerodynamics, fluid dynamics, and other related fields. Initially, two images of fluid flow are taken and divided into frames. Further, these frames are divided into interrogation areas. The cross-correlation is taken between interrogation areas of these two images using discrete cross-correlation (DCC) or by Discrete Fourier Transform using Fast Fourier Transform (FFT). DCC calculates the correlation matrix in the spatial domain. It provides a reliable correlation matrix with low background noise and gives more accurate results compared to standard DFT. The main drawback is it has a high computational cost when compared to DFT. DFT calculates the correlation matrix in the frequency domain and provides high background noise in the correlation matrix with a low computational cost. From the integer displacement of two interrogation areas, the peak intensity of the correlation matrix is computed. The peak detection indicates the particle displacement through sub-pixel accuracy. There may be a certain number of incorrect velocity vectors known as outliers in interrogation areas. To correct these incorrect data, post-processing techniques such as edge detection [15], [16], data validation are used to remove the outliers by choosing the limits of velocity manually. After the removal of outliers, missing data will be replaced by interpolated data. The noise present in the data is removed by the smoothing operation by using the median filter. This paper describes the measurement of vorticity, velocity magnitude, shears, and divergence from the pre-processed image sequence. Techniques like Contrast Limit Adaptive Histogram Equalization (CLAHE) and high pass filter (HPF) are used to enhance (preprocessing) the input image sequence [17].

## II. PROCESS

The PIV measurement technique relies on the physical definition of velocity. The particles in the flow can be captured with CCD or CMOS-camera by illuminating a plane of double pulsed laser in the flow within a time difference of few microseconds. The two sequences of particle images are captured at time t and t'. This allows us to cross-correlate the two-particle images in a small interrogation area to determine their displacement.

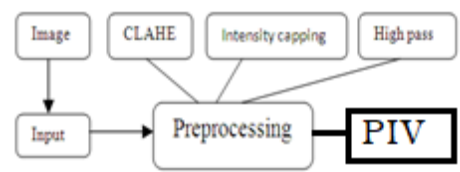

Fig 1 Block diagram of preprocessing technique

### A. Image Pre-processing

Before calculating the correlation of image sequence, enhancement algorithms are applied. CLAHE is one such algorithm that increases the contrast and improves the visibility of the given image. It operates on small regions of the image which resembles like tiles. In each block, the most common intensities of the histogram are ranged out the data [Fig 2.1]. The low-frequency components are eliminated using the HPF. [Fig 2.2].

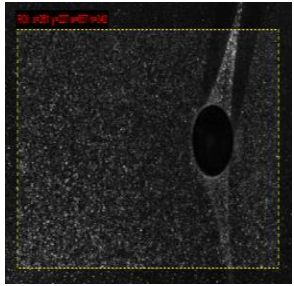 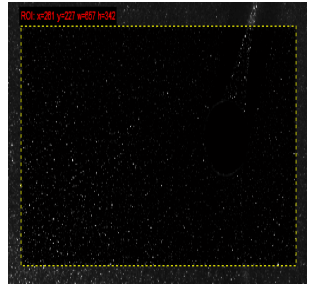

Fig 2.1 Image applied with CLAHE    Fig 2.2 Image applied with HPF

[Fig 2.3] shows the intensity overlaying of the image. In intensity overlaying, pixel intensity is adjusted, which restricts the negative effect of image intensity adjustment.

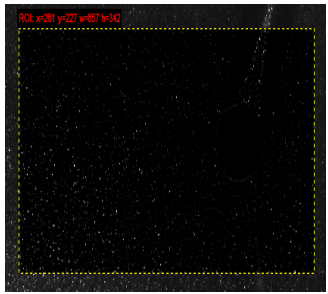

Fig: 2.3 intensity capping

### B. Image Evaluation

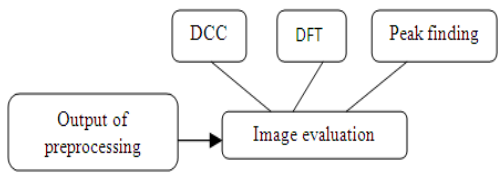

Fig 2.4 Block diagram of image evaluation

After preprocessing, image evaluations of PIV sequences are done. The evaluations include calculating DCC, DFT, and peak finding. The input images are divided into frames and then split into the interrogation area to calculate the correlation takes place using DCC (1) or DFT (2) (3).

$C(m, n) = \sum_i \sum_j A(i, j) B(i + m, j + n)$    - (1)

A and B are examination areas from image engaged at a time t and t'. For calculating DCC, HPF is needed.

$r(i, j) = f(i, j) \otimes g(i, j)$    - (2)

Where r (i, j) is the cross-correlation of the interrogation area of two images f (i, j) and g (i, j). Multiplication of two signals in the time domain is equal to the cross-correlation of that signal in the frequency domain and the inverses also true.

$R(u, v) = F(u, v) \cdot G(u, v)$    - (3)

Where R(u, v) is FFT of r(i, j), F(u, v) and G(u, v) are FFT of f(i, j) and g(i, j).

The displacement of seeded particles in fluid flow is found by finding peaks in cross-correlation [Fig 2.5].

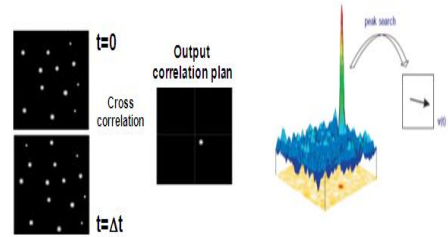

Fig 2.5 Cross-correlation between images taken at t=0 and t=Δ t and finding displacement of particles using peak finding.

### C. Image Post-Processing

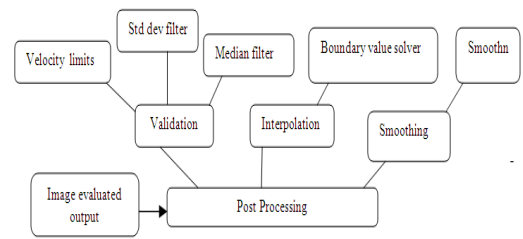

Fig 2.6 Block diagram of image post-processing

The determination of particle displacement may contain invalid data and it can be eliminated by data validation. The correct alignment can be achieved by removing the outliers from the images. Velocity beginnings can also be calculated semi mechanically by relating all velocity components with a lower threshold and an upper threshold ($t_{lower}$ and $t_{upper}$) (4) (5)

$$t_{lower} = \mu - n * \sigma_u \qquad - (4)$$

$$t_{upper} = \mu + n * \sigma_u \qquad - (5)$$

Where $\sigma_u$ is the standard deviation, n is the number of pixels, and µ is the mean. The standard deviation is calculated from the deviation within a neighboring area around each pixel. A standard deviation filter used to highlight the local variability in an image. The removal of incorrect data is replaced by interpolated data using interpolation [Fig 2.7]. There may be a presence of noise that can be removed by smoothing using the median filter. The probability distribution of filter is given by (6),

$$P(X \leq m) \geq \tfrac{1}{2} \quad \text{and} \quad P(X \geq m) \geq \tfrac{1}{2} \qquad - (6)$$

The validation and interpolation steps are to improve the quality of displacement.

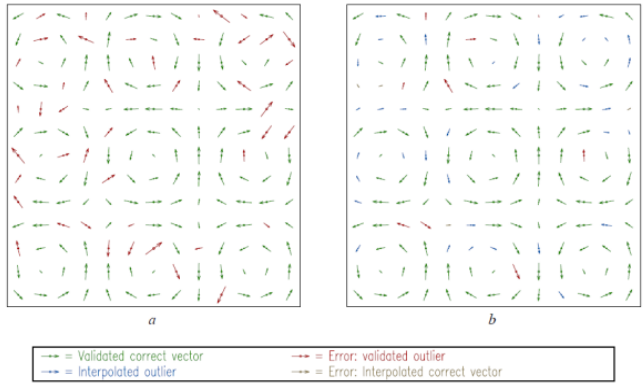

Fig 2.7 a) Vector field containing outliers b) Vector field after validation and interpolation.

### D. Feature Extraction

The existing algorithm determines the different components such as velocity magnitude, shear, strain, divergence, vector, vorticity related to a particular field. In this work, the velocity magnitude and vorticity are computed after the enhancement by CLAHE technique with various clip limits.

*1) Velocity Magnitude:*

Velocity defines the vector quantity that shows the change in distance, time and direction. Velocity shift due to the difference in density is given by,

$$U = d^2_p\,((\rho_p - \rho)/\,18\,\mu)\,a \qquad - (7)$$

Gravitational velocity: a

Where U- velocity, d- displacement, ρ- density

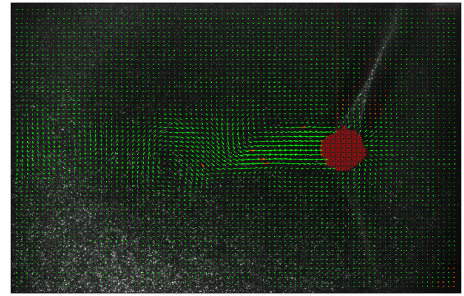

Fig 2.8 Velocity magnitude image

*2) Vorticity:*

The vorticity is a pseudovector field that describes the spinning motion of a continuum which defined as the gradual transition from one condition to another with slight changes near some point understood by an observer positioned at that point and roaming sideways with the movement. Relative to the center of the mass of the particles, vorticity is twice as that of the angular velocity. The vorticity is always perpendicular to the plane of the flow. The vorticity field is defined as the curl of velocity,

$$\omega = \nabla \times u \qquad - (8)$$

$$u = \partial_y \psi - \partial_x \psi \qquad - (9)$$

Where u is the incompressible velocity field, ψ is stream function.

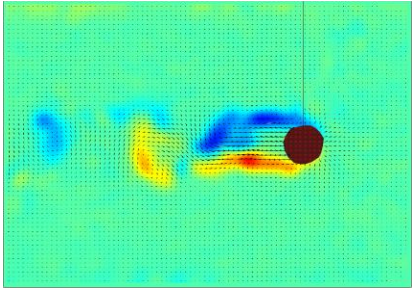

Fig 2.9 Vorticity image

The colors in vorticity denote the levels of the spinning motion of continuum. The red color indicates the highest level of the spinning motion (0.40), the spinning motion of yellow is 0.30, green is 0.20, light blue is 0.10, dark blue is 0.00 its means there is no spinning motion of continuum in the flow of particle.

## III. RESULTS

The experiments are performed on two sequences of images [18] of the fluid particles. It defines the viscous and incompressible motion in the fluid.

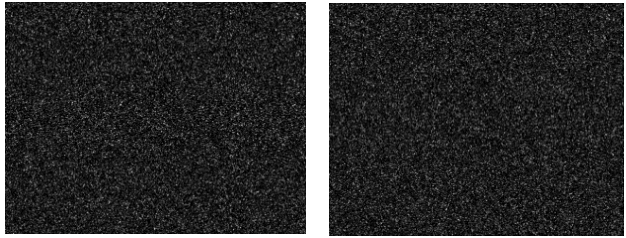

Fig 3.1 a) Image was taken at time t b) Image was taken at time t+ Δt (t').

[Fig 3.2a, b] shows the overlapping of input image and outcomes of the correlation process using FFT and DCC. In DCC, the sample integration area pixel is 24 and the process step is 12. Gaussian 2*3 is used to calculate the sub-pixel images. In FFT, the sample integration area pixel is divided into 4 passes. For passes 1 integration area is 64 passes 2 integration areas is 32, passes 3 integration area is 16, passes 4 integration area is 16. The Linear window deformation method is used to deform the images.

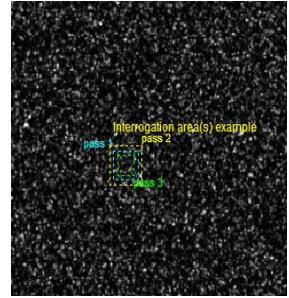

Fig 3.2 a) correlation of image using FFT

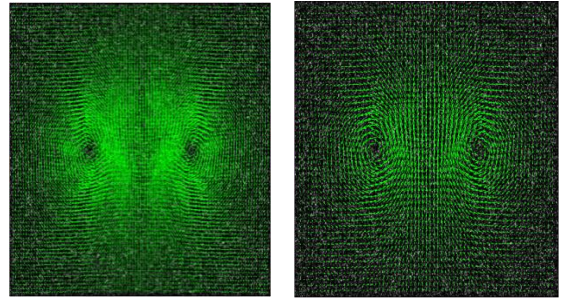

Fig 3.3 a) FFT vector flow image    b) DCC vector flow image

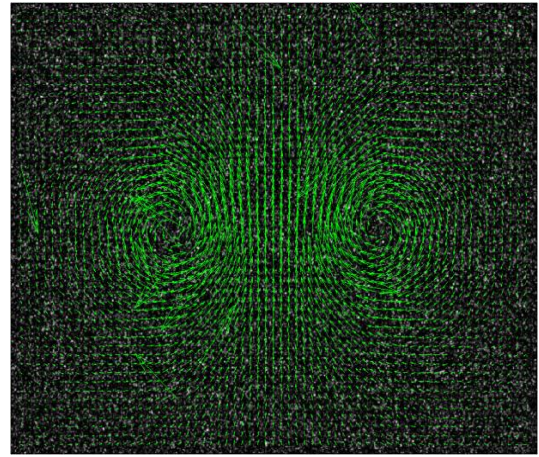

Fig 3.4 Insufficient image pair shows the irrelevant direction of the arrow mark indicates the insufficient image pair of correlation.

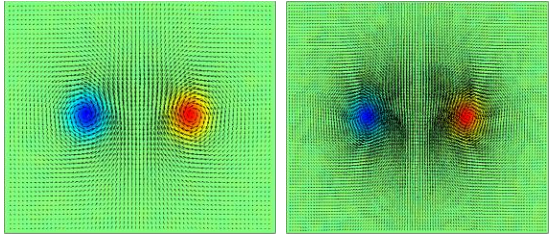

Fig 3.5 a) Vorticity of DCC vector flow image b) Vorticity of FFT vector flow image

The red color in this image shows the high-density of flow and dark blue color indicates the low-density of flow.

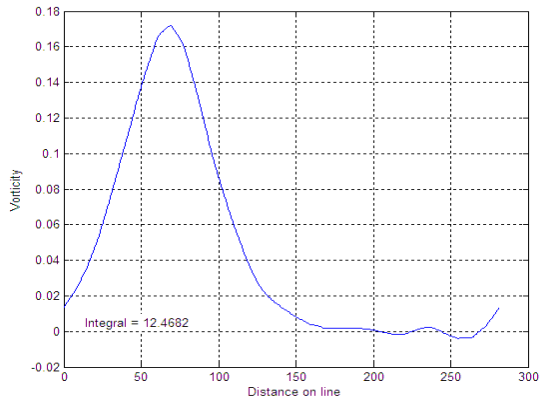

Fig 3.6 Graph of Vorticity Vs distance on line on the high-density flow region.

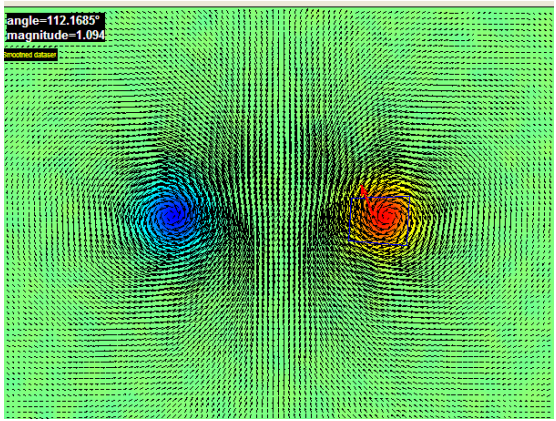

Fig 3.7 Area means flow direction in the highest density flow region.

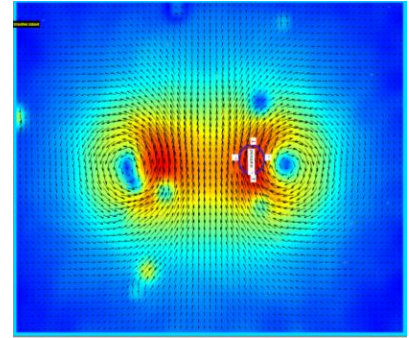

Fig 3.8 Velocity magnitude of the image using DCC

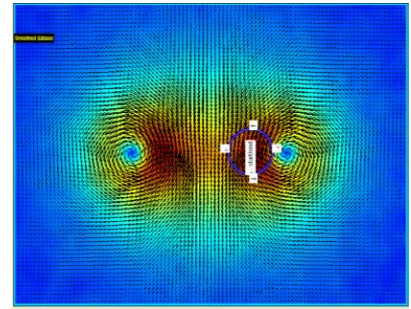

Fig 3.9 Velocity magnitude of the image using FFT

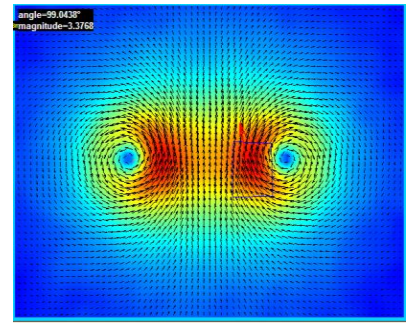

Fig 3.10 Direction of velocity magnitude image

## IV. CONCLUSION

In this work, the enhancement of different types of PIV images was studied, based on the accuracy of velocity, and flow direction was calculated for the given input flow image. It can be used in real-time applications such as swirl combustion, flow in the heart assist system, shear in the river, measurement of flow-induced vibration of hard disk drives, investing pump diffuser flow.